\documentclass[twocolumn,preprintnumbers,nofootinbib,prd]{revtex4}
\usepackage{psfrag,graphicx,epsfig,amsmath,amssymb,bm}
\begin{document}

\preprint{FERMILAB-PUB-09-099-T, MZ-TH/09-11} 

\title{\boldmath
Infrared singularities of QCD amplitudes with massive partons
\unboldmath}

\author{Thomas Becher\,$^a$ and Matthias Neubert\,$^b$} 

\affiliation{$^a$\,Fermi National Accelerator Laboratory,
P.O. Box 500, Batavia, IL 60510, U.S.A.\\
$^b$\,Institut f\"ur Physik (THEP), Johannes Gutenberg-Universit\"at,
D-55099 Mainz, Germany}
\date{\today}

\begin{abstract}
\noindent
A formula for the two-loop infrared singularities of dimensionally regularized QCD scattering amplitudes with an arbitrary number of massive and massless legs is derived. The singularities are obtained from the solution of a renormalization-group equation, and factorization constraints on the relevant anomalous-dimension matrix are analyzed. The simplicity of the structure of the matrix relevant for massless partons does not carry over to the case with massive legs, where starting at two-loop order new color and momentum structures arise, which are not of the color-dipole form. The resulting two-loop three-parton correlations can be expressed in terms of two functions, for which some general properties are derived. This explains observations recently made by Mitov et al.\ in terms of symmetry arguments.
\end{abstract}

\maketitle

\section{Introduction}

In the past few years, remarkable progress has been achieved in the understanding of the infrared (IR) singularities of massless scattering amplitudes in non-abelian gauge theories, which are characterized by an intricate interplay of soft and collinear dynamics. While factorization proofs guarantee the absence of IR divergences in inclusive observables \cite{Collins:1989gx}, an all-order formula for the IR singularities of QCD amplitudes has been lacking for a long time. An important step toward this goal was made by Catani \cite{Catani:1998bh}, who correctly predicted the singularities of two-loop scattering amplitudes apart from the $1/\epsilon$ pole term. An interesting alternative approach to the problem of IR singularities was developed in \cite{Sterman:2002qn}, where the authors exploited the factorization properties of scattering amplitudes along with IR evolution equations to recover Catani's result at two-loop order and relate the coefficient of the $1/\epsilon$ pole term to a soft anomalous-dimension matrix. This matrix was later calculated at two-loop order in \cite{MertAybat:2006wq,MertAybat:2006mz}. Its color structure was found to be of the same form as at one-loop order and it was shown that the more complicated color structure of the $1/\epsilon$-term in Catani's formalism \cite{Bern:2004cz} was an artifact of his subtraction scheme.

In recent work \cite{Becher:2009cu}, we have shown that the IR singularities of on-shell amplitudes in massless QCD are in one-to-one correspondence to the ultra-violet (UV) poles of operator matrix elements in soft-collinear effective theory (SCET) \cite{Bauer:2000yr,Bauer:2001yt,Beneke:2002ph}. They can be subtracted by means of a multiplicative renormalization factor, whose structure is constrained by the renormalization group. We have argued that the simplicity of the corresponding anomalous-dimension matrix holds not only at one- and two-loop order, but is in fact an exact result of perturbation theory. A first test of this prediction at three-loop order was performed in \cite{Dixon:2009gx}. Detailed theoretical arguments supporting our conjecture were presented in \cite{Becher:2009qa}, where we used constraints derived from soft-collinear factorization, the non-abelian exponentiation theorem \cite{Gatheral:1983cz,Frenkel:1984pz}, and the behavior of scattering amplitudes in two-parton collinear limits \cite{Kosower:1999xi} to show that the anomalous-dimension matrix retains its simple form at least to three-loop order, with the possible exception of a single color structure multiplying a function of conformal ratios depending on the momenta of four external partons, which vanishes in all collinear limits. Some of these arguments were developed independently in \cite{Gardi:2009qi}. We also showed that higher Casimir invariants, which in particular would lead to a violation of Casimir scaling of the cusp anomalous dimension \cite{Frenkel:1984pz,Armoni:2006ux,Alday:2007hr,Alday:2007mf}, do not appear at four-loop order. 

It is interesting and relevant for many physical applications to consider generalizations of these results valid in the case of massive partons. The IR singularities of one-loop scattering amplitudes containing massive partons were obtained some time ago in \cite{Catani:2000ef}. We will reproduce their formula as a special case of our analysis. In the limit where the parton masses are small compared with the typical momentum transfer among the partons, they serve as regulators for collinear singularities. In \cite{Penin:2005eh, Mitov:2006xs,Becher:2007cu} factorization theorems for this limit were proposed, which allow one to derive massive amplitudes from massless ones. This technique has been used to derive the massive $e^+ e^- \to e^+ e^-$ scattering amplitude \cite{Penin:2005kf,Becher:2007cu} and the virtual corrections to heavy-quark production in the limit where all kinematic invariants are larger than the heavy-quark masses \cite{Czakon:2007ej,Czakon:2007wk}.
 
First steps toward solving the problem of finding the IR divergences of generic two-loop scattering processes with massive partons, without restricting oneself to the limit of small masses, have recently been taken by Mitov et al.\ \cite{Mitov:2009sv}. Interestingly, these authors find that the simplicity of the anomalous-dimension matrix governing the IR poles does not persist at two-loop order in the massive case. Specifically, they point out that there exist non-vanishing singularities in Feynman graphs connecting three different partons. They evaluate the corresponding contributions numerically and point out that they vanish for some special kinematic configurations.

In this paper we extend our analysis in \cite{Becher:2009qa} to the general case of gauge-theory amplitudes with arbitrary numbers of massive and massless partons. We show that the structure of the terms found in \cite{Mitov:2009sv} follows from simple symmetry arguments, such as soft-collinear factorization in SCET and non-abelian exponentiation. The fact that in the presence of massive partons the low-energy effective theory knows about the 4-velocities of heavy particles in addition to the light-like directions of massless partons gives rise to additional color and kinematical structures, which are absent in the case of massless partons. As a result, the structure of IR poles becomes increasingly more complicated in higher orders of perturbation theory. We present, for the first time, a general formula for the IR singularities of dimensionally regularized two-loop scattering amplitudes with arbitrary numbers of massive and massless partons and arbitrary values of the parton masses. It generalizes the one-loop result of \cite{Catani:2000ef}. For amplitudes with $n\ge 4$ partons, our result contains two new functions with certain symmetry properties, for which at present no analytical expressions are available. 

We begin by considering the case where the parton masses are of the same magnitude as the typical momentum transfer between the partons. In this case the appropriate low-energy effective theory is a combination of SCET and heavy-quark effective theory (HQET) \cite{Neubert:1993mb}, which is applicable since the relative velocities of the heavy partons are of ${\cal O}(1)$ in this case. With the general result at hand, we then explore the limit where the parton masses are taken to be much smaller than the hard momentum transfers between the partons. We conclude that if the two new functions did not vanish in this limit, then the QCD factorization formula of \cite{Mitov:2006xs,Becher:2007cu} would need to be modified.

\section{Soft-collinear factorization}

We denote by $|{\cal M}_n(\epsilon,\{\underline{p}\},\{\underline{m}\})\rangle$, with $\{\underline{p}\}\equiv\{p_1,\dots,p_n\}$ and $\{\underline{m}\}\equiv\{m_1,\dots,m_n\}$, a UV-renormalized, on-shell $n$-parton scattering amplitude with IR singularities regularized in $d=4-2\epsilon$ dimensions. This quantity is a function of the Lorentz invariants $s_{ij}\equiv 2\sigma_{ij}\,p_i\cdot p_j+i0$ and $p_i^2=m_i^2$, where the sign factor $\sigma_{ij}=+1$ if the momenta $p_i$ and $p_j$ are both incoming or outgoing, and $\sigma_{ij}=-1$ otherwise. We assume that all of these invariants are of the same order and refer to them as hard scales. For massive partons ($m_i\ne 0$), we define 4-velocities $v_i=p_i/m_i$, whose components are of ${\cal O}(1)$. We have $v_i^2=1$ and define the abbreviations $w_{ij}\equiv-\sigma_{ij}\,v_i\cdot v_j-i0$. We use the color-space formalism of \cite{Catani:1996jh,Catani:1996vz}, in which $n$-particle amplitudes are treated as $n$-dimensional vectors in color space. $\bm{T}_i$ is the color generator associated with the $i$-th parton and acts as a matrix on its color index. The product $\bm{T}_i\cdot\bm{T}_j\equiv T_i^a\,T_j^a$ is summed over $a$. Generators associated with different particles trivially commute, $\bm{T}_i\cdot\bm{T}_j=\bm{T}_j\cdot\bm{T}_i$ for $i\ne j$, while $\bm{T}_i^2=C_i$ is given in terms of the quadratic Casimir operator of the corresponding color representation, i.e., $C_q=C_{\bar q}=C_F$ for quarks and $C_g=C_A$ for gluons. 

We have shown in \cite{Becher:2009cu,Becher:2009qa} that the IR poles of such amplitudes can be removed by a multiplicative renormalization factor $\bm{Z}^{-1}(\epsilon,\{\underline{p}\},\{\underline{m}\},\mu)$, which acts as a matrix on the color indices of the partons. This quantity obeys the renormalization-group equation 
\begin{equation}\label{RGE}
   \bm{Z}^{-1}\,\frac{d}{d\ln\mu}\,
   \bm{Z}(\epsilon,\{\underline{p}\},\{\underline{m}\},\mu) 
   = - \bm{\Gamma}(\{\underline{p}\},\{\underline{m}\},\mu) \,,
\end{equation}
where $\bm{\Gamma}$ is the anomalous-dimension matrix of effective-theory operators built out of collinear SCET fields for the massless partons and soft HQET fields for the massive ones. The formal solution of this equation is
\begin{equation}\label{Zsolu}
   \bm{Z}(\epsilon,\{\underline{p}\},\{\underline{m}\},\mu) 
   = {\rm\bf{P}} \exp\left[ \int_\mu^{\infty} 
   \frac{{\rm d}\mu'}{\mu'}\,
   \bm{\Gamma}(\{\underline{p}\},\{\underline{m}\},\mu') \right] ,
\end{equation}
where the path-ordering symbol $\rm\bf{P}$ means that matrices are ordered from left to right according to decreasing values of $\mu'$. The $\bm{Z}$-factor appearing in the renormalization of effective-theory operators describes the IR behavior of on-shell amplitudes, because these amplitudes are closely related to the bare Wilson coefficients of the corresponding operators. This connection is discussed in detail in \cite{Becher:2009qa}. Compared to the massless case studied there, we encounter one complication: Since virtual corrections due to heavy quarks are integrated out in the effective theory, the strong coupling constant entering the $\bm{Z}$-factor in the low-energy theory is defined in a theory with massless quark flavors only, while the massive amplitudes in QCD also receive contributions from heavy-quark loops. The $\bm{Z}$-factor we obtain from the effective theory describes the IR singularities of massive QCD amplitudes after the coupling constant is matched onto the effective theory with massless flavors. The corresponding matching relation will be given below.

The interactions between collinear and soft fields can be decoupled by means of a field redefinition \cite{Bauer:2001yt}, after which soft interactions manifest themselves as interactions between a set of light-like and time-like soft Wilson lines representing the massless and massive particles, respectively. Generalizing the discussion of \cite{Becher:2009qa}, the relevant soft operator in the present case is
\begin{equation}\label{Slines}
   {\cal S}(\{\underline{n}\},\{\underline{v}\},\mu) 
   = \langle 0|\bm{S}_{n_1}\ldots\bm{S}_{n_{k}}\,
   \bm{S}_{v_{k+1}}\ldots\bm{S}_{v_n} |0\rangle \,,
\end{equation}
where partons $1,\dots, k$ are massless, and the remaining $n-k$ partons are massive.

From now on, we label the massive partons by capital indices $I,J,\dots$ and the massless ones by lower-case indices $i,j,\dots$. The anomalous-dimension matrix of the effective-theory operators can be written as a sum over soft and collinear contributions \cite{Becher:2009qa},
\begin{equation}\label{decomp}
   \bm{\Gamma}(\{\underline{p}\},\{\underline{m}\},\mu)
   = \bm{\Gamma}_s(\{\underline{\beta}\},\mu) 
   + \sum_i\,\Gamma_c^i(L_i,\mu) \,,
\end{equation}
where $\bm{\Gamma}_s$ is the soft anomalous-dimension matrix governing the UV poles of the Wilson-line operator in (\ref{Slines}). The collinear contributions $\Gamma_c^i$ only arise for massless partons and are diagonal in color space. Note that color conservation implies the relation
\begin{equation}
   \sum_i\,\bm{T}_i + \sum_I\,\bm{T}_I = 0
\end{equation}
when acting on color-singlet states such as color-conserving scattering amplitudes. In intermediate steps in the calculation of the anomalous-dimension matrix one needs to regularize IR divergences in the effective theory, for instance by taking the massless partons slightly off their mass shell, $(-p_i^2)>0$. Following \cite{Becher:2009qa}, we introduce the notation $L_i=\ln[\mu^2/(-p_i^2)]$ for the associated collinear logarithms, which need to cancel in the final result (\ref{decomp}). The soft anomalous-dimension matrix $\bm{\Gamma}_s$ is, in the most general case, a function of the cusp angles $\beta_{ij}$, $\beta_{Ij}$, and $\beta_{IJ}$ formed by the Wilson lines belonging to different pairs of massless or massive partons. With the IR regulator as specified above, the relations expressing these cusp angles in terms of the hard momentum transfers and particle masses read\footnote{Strictly speaking, in the effective theory only the large light-cone components of the collinear momenta appear in the scalar products $p_i\cdot p_j$ and $v_I\cdot p_j$, see \cite{Becher:2009qa}.}
\begin{equation}\label{betasmrel}
\begin{split}
   \beta_{ij} 
   &= \ln\frac{-2\sigma_{ij}\,p_i\cdot p_j\,\mu^2}{(-p_i^2)(-p_j^2)}
    = L_i + L_j - \ln\frac{\mu^2}{-s_{ij}} \,, \\
   \beta_{Ij} 
   &= \ln\frac{-2\sigma_{Ij}\,v_I\cdot p_j\,\mu}{(-p_j^2)}
    = L_j - \ln\frac{m_I\mu}{-s_{Ij}} \,, \\
   \beta_{IJ} 
   &= \mbox{arccosh}(w_{IJ}) 
    = \mbox{arccosh}\Big(\frac{-s_{IJ}}{2m_I m_J}\Big) \,.
\end{split}
\end{equation}
The anomalous dimensions appearing on the right-hand side of (\ref{decomp}) are functions of the cusp angles $\beta_{ij}$ and the collinear logarithms $L_i$, while that on the left-hand side only depends on the hard scales $s_{ij}$ and $m_i$.

\section{Two-parton correlations}

We begin by considering the one- and two-particle terms in the anomalous-dimension matrix (\ref{decomp}). They can be written as \cite{Becher:2003kh}
\begin{equation}\label{Gamc}
   \Gamma_c^i(L_i,\mu) = - C_i\,\gamma_{\rm cusp}(\alpha_s)\,L_i 
   + \gamma_c^i(\alpha_s) \,, 
\end{equation}
and
\begin{equation}\label{Gams}
\begin{split}
   &\bm{\Gamma}_s(\{\underline{\beta}\},\mu)
    \big|_{\rm 2-parton} \\
   &\!= - \!\sum_{(i,j)} \frac{\bm{T}_i\cdot\bm{T}_j}{2}\,
    \gamma_{\rm cusp}(\alpha_s)\,\beta_{ij} 
    - \!\sum_{(I,J)} \frac{\bm{T}_I\cdot\bm{T}_J}{2}\,
    \gamma_{\rm cusp}(\beta_{IJ},\alpha_s) \\
   &\!\quad\mbox{}- \sum_{I,j}\,\bm{T}_I\cdot\bm{T}_j\,
    \gamma_{\rm cusp}(\alpha_s)\,\beta_{Ij} 
    + \sum_i\,\gamma_s^i(\alpha_s)  
    + \sum_I\,\gamma^I(\alpha_s) \,,
\end{split}
\end{equation}
where the notation $(i_1,...,i_k)$ refers to unordered tuples of distinct parton indices. The various coefficients are functions of the renormalized coupling $\alpha_s\equiv\alpha_s(\mu)$ and, in the case of $\gamma_{\rm cusp}(\beta,\alpha_s)$, of a cusp angle $\beta$. The fact that only a linear dependence on the cusp angles is allowed in cases where at least one massless parton is involved has been explained in \cite{Becher:2009qa,Gardi:2009qi}.

It is instructive to see how the dependence on the IR regulators disappears in (\ref{decomp}), when we combine the expressions in (\ref{Gamc}) and (\ref{Gams}) and express the cusp angles in terms of hard momentum transfers and masses as well as collinear logarithms.  We note that (after the cusp angles have been eliminated)
\begin{eqnarray}
   \frac{\partial\bm{\Gamma}_s\big|_{\rm 2-parton}}{\partial L_j}
   &=& - \bigg( \sum_{i\ne j}\,\bm{T}_i\cdot\bm{T}_j
    + \sum_I\,\bm{T}_I\cdot\bm{T}_j \bigg)\, 
    \gamma_{\rm cusp}(\alpha_s) \nonumber\\
   &=& C_j\,\gamma_{\rm cusp}(\alpha_s) 
    = - \frac{\partial\Gamma_c^j}{\partial L_j} \,.
\end{eqnarray}
Hence, the sum of all contributions is indeed independent of the IR regulators. Note that this requirement fixes the relative strength of the terms proportional to $\bm{T}_i\cdot\bm{T}_j$ and $\bm{T}_I\cdot\bm{T}_j$ in (\ref{Gams}). From (\ref{decomp}) we then obtain
\begin{equation}\label{resu1}
\begin{split}
   &\bm{\Gamma}(\{\underline{p}\},\{\underline{m}\},\mu)
    \big|_{\rm 2-parton}\\
   &= \sum_{(i,j)}\,\frac{\bm{T}_i\cdot\bm{T}_j}{2}\,
    \gamma_{\rm cusp}(\alpha_s)\,\ln\frac{\mu^2}{-s_{ij}}
    + \sum_i\,\gamma^i(\alpha_s) \\
   &\quad\mbox{}- \sum_{(I,J)}\,\frac{\bm{T}_I\cdot\bm{T}_J}{2}\,
    \gamma_{\rm cusp}(\beta_{IJ},\alpha_s)
    + \sum_I\,\gamma^I(\alpha_s) \\
   &\quad\mbox{}+ \sum_{I,j}\,\bm{T}_I\cdot\bm{T}_j\,
    \gamma_{\rm cusp}(\alpha_s)\,\ln\frac{m_I\mu}{-s_{Ij}} \,,
\end{split}
\end{equation}
where $\gamma^i\equiv\gamma_s^i+\gamma_c^i$. 

The anomalous-dimension coefficients $\gamma_{\rm cusp}(\alpha_s)$ and $\gamma^i(\alpha_s)$ (for $i=q,g$) have been determined to three-loop order in \cite{Becher:2009qa} by considering the case of the massless quark and gluon form factors. For example, the one- and two-loop coefficients in the perturbative series $\gamma_{\rm cusp}(\alpha_s)=\sum_n\gamma_n^{\rm cusp}\,(\frac{\alpha_s}{4\pi})^{n+1}$ are
\begin{equation}
   \gamma_0^{\rm cusp} = 4 \,, \qquad
   \gamma_1^{\rm cusp} = \left( \frac{268}{9} - \frac{4\pi^2}{3}
    \right) C_A - \frac{80}{9}\,T_F n_f \,.
\end{equation}
In QCD only quarks can be massive, and the first two coefficients in the expansion of $\gamma^Q$ can be extracted by matching our result with the known form of the anomalous dimension of heavy-light currents $J_{\rm hl}$ in SCET \cite{Bauer:2001yt,Bosch:2003fc}. We obtain
\begin{equation}
   \Gamma_{J_{\rm hl}}(p,v,\mu) = - C_F\,\gamma_{\rm cusp}(\alpha_s)\,
   \ln\frac{\mu}{2v\cdot p} + \gamma'(\alpha_s) \,,
\end{equation}
where we assume that the heavy-quark with velocity $v$ is incoming and the light quark with momentum $p$ is outgoing. The sum $\gamma'=\gamma^q+\gamma^Q$ was first obtained at two-loop order in \cite{Neubert:2004dd}. Using this result leads to the one- and two-loop coefficients
\begin{equation}\label{gammaQ}
\begin{split}
   \gamma_0^Q &= - 2 C_F \,, \\
   \gamma_1^Q &= C_F C_A \left( \frac{2\pi^2}{3} - \frac{98}{9} 
    - 4\zeta_3 \right) + \frac{40}{9}\,C_F T_F n_f \,.
\end{split}
\end{equation}
Since this anomalous dimension is connected with the soft Wilson-line operator in (\ref{Slines}), its color structures are constrained by the non-abelian exponentiation theorem \cite{Gatheral:1983cz,Frenkel:1984pz}. This explains the absence of a $C_F^2$ term, and it implies that up to ${\cal O}(\alpha_s^3)$ the corresponding anomalous dimension for a massive color-octet particle is given by $C_A/C_F$ times the coefficients shown above.

It remains to determine the velocity-dependent function $\gamma_{\rm cusp}(\beta,\alpha_s)$ in (\ref{resu1}). This can be accomplished by applying our general result to the special case of just two heavy quarks, where it should reproduce the velocity-dependent anomalous dimension of heavy-quark currents in HQET \cite{Neubert:1993mb,Falk:1990yz}. It follows that
\begin{equation}
   \Gamma_{J_{\rm hh}}(v,v',\alpha_s)
   = C_F\,\gamma_{\rm cusp}(\beta,\alpha_s) + 2\gamma^Q(\alpha_s)
\end{equation}
with $\cosh\beta=v\cdot v'$, where $v$ is the incoming and $v'$ the outgoing quark's velocity. This anomalous dimension was calculated at two-loop order in \cite{Korchemsky:1987wg,Korchemsky:1991zp}, and it has recently been recomputed and expressed in compact form in \cite{Kidonakis:2009ev}. Using the result (\ref{gammaQ}) for $\gamma^Q$, we obtain from the latter paper
\begin{widetext}
\begin{equation}
\begin{split}
   \gamma_{\rm cusp}(\beta,\alpha_s)
   &= \gamma_{\rm cusp}(\alpha_s)\,\beta\coth\beta 
    + \frac{C_A}{2} \left( \frac{\alpha_s}{\pi} \right)^2 
    \Bigg\{ \coth^2\beta \left[ \mbox{Li}_3(e^{-2\beta}) 
   + \beta\,\mbox{Li}_2(e^{-2\beta}) - \zeta_3 
   + \frac{\pi^2}{6}\,\beta + \frac{\beta^3}{3} \right] \\
  &\quad\mbox{}+ \coth\beta \left[ 
   \mbox{Li}_2(e^{-2\beta}) - 2\beta\,\ln(1-e^{-2\beta}) 
   - \frac{\pi^2}{6}\,(1+\beta) - \beta^2 - \frac{\beta^3}{3} 
   \right] + \frac{\pi^2}{6} + \zeta_3 + \beta^2 \Bigg\}
   + \dots \,.
\end{split}
\end{equation}
\end{widetext}
For small cusp angle the result can be expanded in even powers of $\beta$, 
\begin{eqnarray}
   \gamma_{\rm cusp}(\beta,\alpha_s) 
   &=& \gamma_{\rm cusp}(\alpha_s) \left( 1 + \frac{\beta^2}{3}
    + \dots \right) \\
   &&\hspace{-1.1cm}\mbox{}+ C_A \left( \frac{\alpha_s}{\pi} \right)^2
    \left[ \frac{\zeta_3-1}{2}
    + \left( \frac14 - \frac{\pi^2}{36} \right) \beta^2 
    + \dots \right] . \nonumber
\end{eqnarray}
Note that the leading terms in the expansion of $\gamma_{\rm cusp}(\beta,\alpha_s)$ around small cusp angle are equal to $-2\gamma_Q$, so that the full anomalous dimension $\Gamma_{J_{\rm hh}}(v,v',\alpha_s)$ vanishes in the limit $v \cdot v'=1$.
In the limit of large cusp angle one finds
\begin{equation}
   \gamma_{\rm cusp}(\beta,\alpha_s)
   = \gamma_{\rm cusp}(\alpha_s)\,\beta + \dots \,,
\end{equation}
where the dots represent terms that vanish for $\beta\to\infty$. Note that no constant terms remain in this limit.

The $\bm{Z}$-factor associated with a renormalization-group equation such as (\ref{RGE}), in which the anomalous dimension is linear in $\ln\mu$, was derived in \cite{Becher:2009cu}. To two-loop order it reads
\begin{eqnarray}\label{result}
   \ln\bm{Z} 
   &=& \frac{\alpha_s}{4\pi} 
    \left( \frac{\Gamma_0'}{4\epsilon^2}
    + \frac{\bm{\Gamma}_0}{2\epsilon} \right) \\
   &&\mbox{}+ \left( \frac{\alpha_s}{4\pi} \right)^2 \! 
    \left[ - \frac{3\beta_0\Gamma_0'}{16\epsilon^3} 
    + \frac{\Gamma_1'-4\beta_0\bm{\Gamma}_0}{16\epsilon^2}
    + \frac{\bm{\Gamma}_1}{4\epsilon} \right] + \dots \,, \nonumber
\end{eqnarray}
 where 
\begin{equation}
   \Gamma'(\alpha_s) 
   = \frac{\partial}{\partial\ln\mu}\,
    \bm{\Gamma}(\{\underline{p}\},\{\underline{m}\},\mu)
   = - \gamma_{\rm cusp}(\alpha_s)\,\sum_i\,C_i \,,
\end{equation}
and we have expanded $\bm{\Gamma}=\bm{\Gamma}_0\,\frac{\alpha_s}{4\pi}+\bm{\Gamma}_1\,(\frac{\alpha_s}{4\pi})^2+\dots$. Exponentiating the result (\ref{result}) yields the two-loop expression for $\bm{Z}$, which encodes the IR singularities of the massive QCD scattering amplitudes. More precisely, we have
\begin{equation}
   \bm{Z}^{-1}(\alpha_s)\,
   |{\cal M}_n(\epsilon,\{\underline{p}\},\{\underline{m}\})\rangle
   \big|_{\alpha_s^{\rm QCD}\to\xi\alpha_s} = \mbox{finite}
\end{equation}
for $\epsilon\to 0$. The quantity $\alpha_s$ denotes the strong coupling constant in the effective theory, which is obtained after integrating out the heavy quark flavors. It is obtained from the coupling constant $\alpha_s^{\rm QCD}$ of full QCD via the decoupling relation $\alpha_s^{\rm QCD}=\xi\alpha_s$. To first order in $\alpha_s$, the matching factor appropriate for $n_h$ heavy-quark flavors reads \cite{Steinhauser:2002rq}
\begin{equation}\label{eq:decouple}
   \xi = 1 + \frac{\alpha_s}{3\pi}\,T_F\,\sum_{i=1}^{n_h} 
   \left[ e^{\epsilon\gamma_E}\,\Gamma(\epsilon) 
   \left( \frac{\mu^2}{m_i^2} \right)^\epsilon - \frac{1}{\epsilon} 
   \right] .
\end{equation}
We have checked that at one-loop order our results (\ref{resu1}) and  (\ref{result}) reproduce the IR pole terms obtained in \cite{Catani:2000ef}. We have also confirmed that they correctly describe the IR singularities of the two-loop massive quark form factor calculated in \cite{Bernreuther:2004ih,Bernreuther:2005gq,Bernreuther:2005rw}.

Our result (\ref{resu1}) has been derived under the assumption of hard parton masses, in which case the appropriate low-energy effective theory is well known. However, being an exact result in perturbative QCD, this formula can also be applied in cases where some or all of the parton masses are much smaller than the momentum transfers between the partons, $m_I m_J\ll|s_{IJ}|$. This limit is relevant to many processes, such as Bhabha scattering or  heavy-quark production at the LHC. It is the limit in which all cusp angles, also those involving massive partons, become large, and $\beta_{IJ}=\ln(-s_{IJ}/m_I m_J)$ up to power-suppressed terms. In this case our general result (\ref{resu1}) implies that
\begin{equation}
\begin{split}
   &\bm{\Gamma}(\{\underline{p}\},\{\underline{m}\to 0\},\mu)
    \big|_{\rm 2-parton} 
    - \bm{\Gamma}(\{\underline{p}\},\{\underline{0}\},\mu) \\
   &= \sum_I \left[
    C_I\,\gamma_{\rm cusp}(\alpha_s)\,\ln\frac{\mu}{m_I}
    + \gamma^I(\alpha_s) - \gamma^i(\alpha_s) \right] ,
\end{split}
\end{equation}
where $\bm{\Gamma}(\{\underline{p}\},\{\underline{0}\},\mu)$ is the anomalous-dimension matrix in the massless case, whose conjectured all-order form is given by the terms shown in the second line in (\ref{resu1}) \cite{Becher:2009cu,Becher:2009qa}. In the equation above, $\gamma^i$ is the massless single-parton anomalous dimension belonging to parton $I$. In QCD only quarks can be massive, and this result can be rewritten as a sum over heavy-quark anomalous dimensions
\begin{equation}\label{gammaQeq}
   \Gamma_Q(m_Q,\mu) 
   = C_F\,\gamma_{\rm cusp}(\alpha_s)\,\ln\frac{\mu}{m_Q}
   + \gamma^Q(\alpha_s) - \gamma^q(\alpha_s) \,,
\end{equation} 
where the one- and two-loop coefficients of the constant terms are
\begin{equation}
\begin{split}
   \gamma_0^Q - \gamma_0^q &= C_F \,, \\
   \gamma_1^Q - \gamma_1^q 
   &= C_F^2 \left( \frac32 - 2\pi^2 + 24\zeta_3 \right) \\
   &\quad\mbox{}+ C_F C_A \left( \frac{373}{54} + \frac{5\pi^2}{2} 
    - 30\zeta_3 \right) \\
   &\quad\mbox{}- C_F T_F n_f \left( \frac{10}{27}
    + \frac{2\pi^2}{3} \right) .
\end{split}
\end{equation}
The factor $Z_Q$ associated with (\ref{gammaQeq}), which is obtained after substituting the anomalous dimension $\Gamma_Q$ into the general relation (\ref{result}), is compatible with the results of \cite{Mitov:2006xs,Becher:2007cu}. Specifically, we find that the product $Z_Q^{-2} \,Z^{\{m|0\}}$ is finite, where the quantity $Z^{\{m|0\}}$ was defined in \cite{Mitov:2006xs} as the ratio of the massive to the massless quark form factor in the limit where the quark mass tends to zero, and without including heavy fermion loops. Note that 
our derivation assumed that the massive partons are heavy enough to be integrated out in the low-energy theory using (\ref{eq:decouple}). If this is not the case, then the treatment of the heavy-flavor contribution is more complicated \cite{Becher:2007cu}.

\section{Three-parton correlations}
\label{sec:3parton}

It was observed in \cite{Mitov:2009sv} that in the case with massive partons the anomalous-dimension matrix (\ref{decomp}) has a more complicated structure than in the massless case, and that at two-loop order non-abelian diagrams connecting three partons give rise to non-vanishing contributions. The additional terms were found to vanish if two of the three partons are massless,\footnote{It is noted in \cite{Mitov:2009sv} that this observation has been made independently by Einan Gardi.} 
or if any pair of the three kinematic invariants formed out of the parton momenta are equal. We will now show that these observations have a simple explanation.

\begin{figure}
\begin{center}
\psfrag{i}{$n_i$}\psfrag{j}{$n_j$}\psfrag{k}{$n_k$}
\psfrag{I}[r]{$v_I$}\psfrag{J}{$v_J$}\psfrag{K}{$v_K$}
\psfrag{d}{$\vdots$}
\includegraphics[width=0.42\columnwidth]{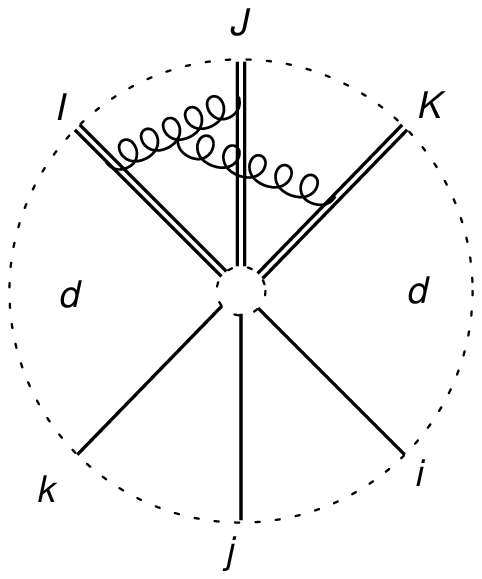}\hspace{0.05\columnwidth}\includegraphics[width=0.42\columnwidth]{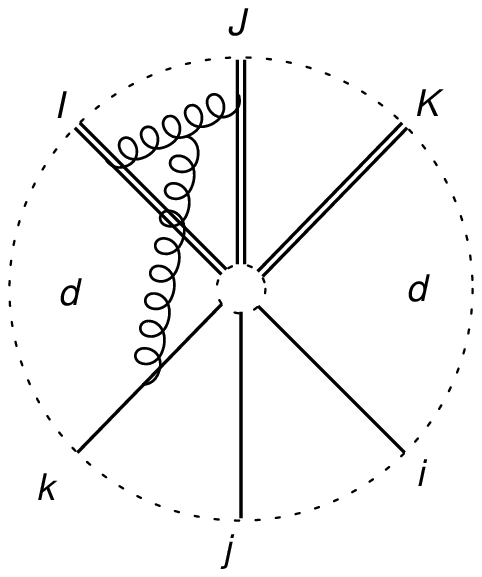} 
\end{center}
\vspace{-4mm}
\caption{\label{fig:graphs}
Graphical representation of the two three-particle terms in the anomalous-dimension matrix (\ref{resu2}). Double lines represent massive partons, single lines show massless ones.}
\end{figure}

Adapting the diagrammatic analysis of our paper \cite{Becher:2009qa} to the case with non-zero parton masses, we find that additional structures arise from two-loop order on, the reason being that the 4-velocities of the massive partons are known to both the full and the effective theories. In HQET the velocities appear as labels on the effective heavy-quark fields \cite{Neubert:1993mb,Georgi:1990um}. In the full theory, they are simply given by $v_i=p_i/m_i$. While for massless partons the rewriting from hard to soft variables always introduces collinear logarithms, this is not true for massive partons, as shown in (\ref{betasmrel}). At two-loop order, the non-abelian exponentiation theorem then allows additional structures involving three partons. They are absent in the massless case, because it is impossible to form a totally anti-symmetric function of three cusp angles $\beta_{ij}$, $\beta_{jk}$, $\beta_{ki}$ that is independent of collinear logarithms upon the substitution shown in the first line in (\ref{betasmrel}) \cite{Becher:2009qa}. This would violate soft-collinear factorization. However, with massive partons this argument no longer applies. In fact, in principle the soft anomalous-dimension matrix can contain the structures 
\begin{equation}\label{F3}
\begin{split}
   &\bm{\Gamma}_s(\{\underline{\beta}\},\mu)
    \big|_{\rm 3-parton} \\
   &= i f^{abc} \sum_{(I,J,K)} \bm{T}_I^a\,\bm{T}_J^b\,\bm{T}_K^c\,
    F_1(\beta_{IJ},\beta_{JK},\beta_{KI}) \\
   &\quad\mbox{}+ i f^{abc} \sum_{(I,J)} \sum_k\,
    \bm{T}_I^a\,\bm{T}_J^b\,\bm{T}_k^c\,
    F_2(\beta_{IJ},\beta_{Jk},\beta_{Ik}) \\
   &\quad\mbox{}+ i f^{abc} \sum_I \sum_{(j,k)}\,
    \bm{T}_I^a\,\bm{T}_j^b\,\bm{T}_k^c\,
    F_3(\beta_{Ij},\beta_{Ik},\beta_{jk}) \,.
\end{split}
\end{equation}
The function $F_1$ must be totally anti-symmetric in its arguments, while $F_2$ ($F_3$) must be anti-symmetric in the last (first) two arguments. Soft-collinear factorization enforces that after elimination of the cusp angles using (\ref{betasmrel}) the result (\ref{F3}) must be independent of collinear logarithms. This in turn requires that
\begin{equation}
\begin{split}
   F_2(\beta_{IJ},\beta_{Jk},\beta_{Ik}) 
   &= f_2(\beta_{IJ},\beta_{Jk}-\beta_{Ik}) \,, \\
   F_3(\beta_{Ij},\beta_{Ik},\beta_{jk}) 
   &= 0 \,,
\end{split}
\end{equation}
where $f_2(x,y)$ must be an odd function of $y$. Note that for $F_3$ to be independent of collinear logarithms it should be a function of the combination $(\beta_{Ij}+\beta_{Ik}-\beta_{jk})$, but this is symmetric in $j,k$ and so vanishes when contracted with the anti-symmetric color structure. Hence only the two possibilities illustrated in Figure~\ref{fig:graphs} remain, and we are led to the following additional structures in the complete anomalous-dimension matrix in (\ref{decomp}):
\begin{eqnarray}\label{resu2}
   &&\hspace{-4mm}\bm{\Gamma}(\{\underline{p}\},\{\underline{m}\},\mu)
    \big|_{\rm 3-parton} \nonumber\\
   &=& i f^{abc} \sum_{(I,J,K)} \bm{T}_I^a\,\bm{T}_J^b\,\bm{T}_K^c\,
    F_1(\beta_{IJ},\beta_{JK},\beta_{KI}) \\
   &&\mbox{}+ i f^{abc} \sum_{(I,J)} \sum_k\,
    \bm{T}_I^a\,\bm{T}_J^b\,\bm{T}_k^c\,
    f_2\Big(\beta_{IJ},
     \ln\frac{-\sigma_{Jk}\,v_J\cdot p_k}%
             {-\sigma_{Ik}\,v_I\cdot p_k}\Big) \,. \nonumber
\end{eqnarray}
This is the most general form possible at two-loop order, and hence the sum of (\ref{resu1}) and (\ref{resu2}) gives the complete answer for the anomalous-dimension matrix at ${\cal O}(\alpha_s^2)$. We note that the color factor in (\ref{resu2}) is non-zero only if there are at least four partons involved in the scattering process. For three partons, charge conservation, i.e. the fact that $\bm{T}_1+\bm{T}_2+\bm{T}_3=0$, implies that
\begin{equation}
   f^{abc}\,\bm{T}_1^a\,\bm{T}_2^b\,\bm{T}_3^c
   = - f^{abc}\,\bm{T}_1^a\,\bm{T}_2^b\,(\bm{T}_1^c + \bm{T}_2^c)
   = 0 \,.
\end{equation}
Since the color matrices entering the two-particle and three-particle terms do not commute, the path ordering in (\ref{Zsolu}) becomes important in the massive case. However, since the three-parton color structures first enter at two-loop order, this complication arises only at ${\cal O}(\alpha_s^3)$.

It is interesting to observe that the two structures in (\ref{resu2}) are consistent with the constraints following from the behavior of scattering amplitudes in the limit where two (or more) massless partons become collinear.  To see this, consider the case where the first two light partons become collinear, such that $p_1=z P$ and $p_2=(1-z) P$ with $P^2\to 0$. As shown in \cite{Becher:2009qa}, the anomalous dimension of the matrix of splitting amplitudes $\mbox{\bf Sp}(\{p_1,p_2\},\mu)$ is given by the difference of $\bm{\Gamma}(\{p_1,p_2,p_3,\dots\},\{\underline{m}\},\mu)$ and $\bm{\Gamma}(\{P,p_3,\dots\},\{\underline{m}\},\mu)|_{\bm{T}_P=\bm{T}_1+\bm{T}_2}$, and it must be independent of the momenta and colors of the remaining partons. Indeed, the three-parton term (\ref{resu2}) does not contribute to this difference, because it is at most linear in the color generators of massless partons and invariant under rescalings of their momenta. It would be interesting to explore whether additional constraints on the anomalous dimension arise from considering quasi-collinear limits \cite{Catani:2000ef,Catani:2002hc}, a generalization of the usual collinear limit to the case where some of the partons involved in the splitting are massive. 

Our result (\ref{resu2}) explains the observation made in \cite{Mitov:2009sv}, that the three-parton correlations do not arise when two or three of the involved partons are massless (i.e., the fact that $F_3=0$). Moreover, when any pair of kinematic invariants are equal, these terms vanish as well. For $F_1$ this follows from the fact that it is an anti-symmetric function in all of its arguments. For $f_2$ it follows since for $v_I=v_J$ both arguments of the function vanish, but $f_2$ must be odd in its second argument. We can thus reproduce and understand all of the observations made in \cite{Mitov:2009sv} based on symmetry properties and without an explicit two-loop calculation. This also demonstrates that these observations will continue to hold in higher orders of perturbation theory. On the other hand, due to reasons similar to the ones outlined above, it is clear that when massive particles are present more and more complicated color and momentum structures will arise in higher orders of perturbation theory. The arguments of \cite{Becher:2009qa} show that all of these structures are of non-abelian origin and involve three or more partons.

An integral representation for the function $F_1$ can, in principle, be extracted from expressions derived in \cite{Mitov:2009sv}. These authors have shown by numerical evaluation that $F_1\ne 0$ for generic values of its arguments. It is an open question, however, how the functions $F_1$ and $f_2$ behave in the limit of large cusp angles, corresponding to the case where $m_I m_J\ll|s_{IJ}|$. If either one of the two functions does not vanish in this limit, then the factorization theorem for massive amplitudes proposed in \cite{Mitov:2006xs,Becher:2007cu} would have to be modified to account for the non-factorizable three-particle terms derived from (\ref{resu2}). For example, in heavy-quark production processes such as $q\bar q\to t\bar t$ or $gg\to t\bar t$ only the second term in (\ref{resu2}) contributes. If the asymptotic behavior of the coefficient function were $f_2(x,y)\sim x y$, then in the limit $s\gg m_t^2$ the three-parton term would contribute a $1/\epsilon$ IR pole at two-loop order proportional to $\ln(t/u) \ln(-s/m_t^2)$, which would be incompatible with a simple factorization formula. The agreement of the numerical results presented in \cite{Czakon:2008zk} for the $q\bar q\to t\bar t$ scattering amplitude with the predictions obtained in \cite{Czakon:2007ej} using the factorization theorem for massive partons provides some evidence that the three-parton contributions do indeed vanish in the limit of large cusp angles, but one should check this with an explicit calculation of $F_1$ and $f_2$.

\section{Summary and outlook}

We have derived the general form of infrared singularities of two-loop QCD amplitudes involving arbitrary numbers of massive and massless partons. These singularities can be absorbed into a multiplicative $\bm{Z}$-factor, which fulfills a renormalization-group equation.
In the purely massless case, the associated anomalous dimension involves only color-dipole correlations. This was first observed by explicit calculation at the two-loop level in \cite{MertAybat:2006wq,MertAybat:2006mz}. In a recent paper, we have argued that factorization constraints suggest that this property persists to all orders and have analyzed the three-loop case in detail \cite{Becher:2009qa}. More recently, Mitov et al.\ \cite{Mitov:2009sv} have shown that the anomalous-dimension matrix for massive partons involves color correlations among three partons, in contrast to the massless case. They have evaluated these contributions numerically in Euclidean space and pointed out that they vanish for certain kinematical configurations.

In the present paper, we have derived the structure of the two-loop anomalous-dimension matrix relevant for the general case involving both massive and massless partons. The factorization constraints can be studied using a combination of soft-collinear effective theory for the massless partons and heavy-quark effective theory for the massive ones. We have first derived the form of the two-parton correlations and extracted the necessary anomalous dimensions from known results for heavy-to-heavy and heavy-to-light currents in the effective theory. The factorization constraints are weaker for terms involving massive legs, since both the full and the effective theories know about the four-velocities of the massive partons. In particular, the constraints do not exclude three-parton correlations if at least two of the partons involved are massive. We have shown that two such structures appear and that their symmetry properties imply that they vanish when two four-velocities of the involved massive partons become equal, which explains the findings of \cite{Mitov:2009sv}. Starting from their results, one should derive explicit representations for these two functions in Minkowski space, preferably in analytic form.

We have briefly discussed the limit in which the particle masses are small compared to the momentum transfers. In this case, the two-parton contribution factors into a hard part, depending on the large momentum transfers, and a sum over collinear contributions depending on the parton masses. It would be interesting and important to study this case in more detail. In particular, one should check whether the three-parton correlations vanish in this limit, since their presence would violate the factorization theorem proposed in \cite{Mitov:2006xs, Becher:2007cu}.

{\em Acknowledgments:\/}
M.N.~is grateful to Uli Haisch and Andrea Ferroglia for useful discussions. The research of T.B.\ was supported by the U.S.\ Department of Energy under Grant DE-AC02-76CH03000. Fermilab is operated by the Fermi Research Alliance under contract with the Department of Energy.

\section*{Note added}

Soon after this paper was published, the two-loop expressions for the functions $F_1$ and $f_2$ describing the three-parton correlations in (\ref{resu2}) have been calculated in closed analytic form \cite{Ferroglia:2009ep,Ferroglia:2009ii}. In light of these results, some statements made in the last two paragraphs of Section~\ref{sec:3parton} need to be revised:

1.~The conclusion drawn in \cite{Mitov:2009sv} and in the present work, that the three-parton terms vanish whenever the four-velocities of two massive partons coincide, is false. Due to the presence of Coulomb singularities the limit $v_I\to v_J$ is singular. The three-parton terms do not vanish but diverge logarithmically in this limit.

2.~A formal argument presented in \cite{Mitov:2009sv}, suggesting that the three-parton terms only receive contributions from the three-gluon graphs shown in Figure~\ref{fig:graphs}, is invalid. Additional contributions with the same color structure arise from two-loop planar and one-loop counterterm diagrams. The calculations presented in that paper are therefore not sufficient to extract an integral representation for the function $F_1$.

We also note that the explicit expressions for the functions $F_1$ and $f_2$ vanish like $(m_I m_J/s_{IJ})^2$ in the limit of small parton masses. The factorization theorem for amplitudes with massive partons proposed in \cite{Mitov:2006xs,Becher:2007cu} is therefore not invalidated by their presence.


\begin{thebibliography}{99}

\bibitem{Collins:1989gx}
  J.~C.~Collins, D.~E.~Soper and G.~Sterman,
  Adv.\ Ser.\ Direct.\ High Energy Phys.\  {\bf 5}, 1 (1988)
  [arXiv:hep-ph/0409313].
    
\bibitem{Catani:1998bh}
  S.~Catani,
  Phys.\ Lett.\  B {\bf 427}, 161 (1998)
  [arXiv:hep-ph/9802439].

\bibitem{Sterman:2002qn}
  G.~Sterman and M.~E.~Tejeda-Yeomans,
  Phys.\ Lett.\  B {\bf 552}, 48 (2003)
  [arXiv:hep-ph/0210130].

\bibitem{MertAybat:2006wq}
  S.~M.~Aybat, L.~J.~Dixon and G.~Sterman,
  Phys.\ Rev.\ Lett.\  {\bf 97}, 072001 (2006) 
  [arXiv:hep-ph/0606254].

\bibitem{MertAybat:2006mz}
  S.~M.~Aybat, L.~J.~Dixon and G.~Sterman,
  Phys.\ Rev.\  D {\bf 74}, 074004 (2006)
  [arXiv:hep-ph/0607309].
   
\bibitem{Bern:2004cz}
  Z.~Bern, L.~J.~Dixon and D.~A.~Kosower,
  JHEP {\bf 0408}, 012 (2004)
  [arXiv:hep-ph/0404293].

\bibitem{Becher:2009cu}
  T.~Becher and M.~Neubert,
  Phys.\ Rev.\ Lett.\  {\bf 102}, 162001 (2009)
  [arXiv:0901.0722 [hep-ph]].

\bibitem{Bauer:2000yr}
  C.~W.~Bauer, S.~Fleming, D.~Pirjol and I.~W.~Stewart,
  Phys.\ Rev.\  D {\bf 63}, 114020 (2001)
  [arXiv:hep-ph/0011336].

\bibitem{Bauer:2001yt}
  C.~W.~Bauer, D.~Pirjol and I.~W.~Stewart,
  Phys.\ Rev.\  D {\bf 65}, 054022 (2002)
  [arXiv:hep-ph/0109045].

\bibitem{Beneke:2002ph}
  M.~Beneke, A.~P.~Chapovsky, M.~Diehl and T.~Feldmann,
  Nucl.\ Phys.\  B {\bf 643}, 431 (2002)
  [arXiv:hep-ph/0206152].

\bibitem{Dixon:2009gx}
  L.~J.~Dixon,
  arXiv:0901.3414 [hep-ph].

\bibitem{Becher:2009qa}
  T.~Becher and M.~Neubert,
  JHEP {\bf 0906}, 081 (2009)
  [arXiv:0903.1126 [hep-ph]].

\bibitem{Gatheral:1983cz}
  J.~G.~M.~Gatheral,
  Phys.\ Lett.\  B {\bf 133}, 90 (1983).

\bibitem{Frenkel:1984pz}
  J.~Frenkel and J.~C.~Taylor,
  Nucl.\ Phys.\  B {\bf 246}, 231 (1984).

\bibitem{Kosower:1999xi}
  D.~A.~Kosower,
  Nucl.\ Phys.\  B {\bf 552}, 319 (1999)
  [arXiv:hep-ph/9901201].

\bibitem{Gardi:2009qi}
  E.~Gardi and L.~Magnea,
  JHEP {\bf 0903}, 079 (2009)
  [arXiv:0901.1091 [hep-ph]].

\bibitem{Armoni:2006ux}
  A.~Armoni,
  JHEP {\bf 0611}, 009 (2006)
  [arXiv:hep-th/0608026].

\bibitem{Alday:2007hr}
  L.~F.~Alday and J.~M.~Maldacena,
  JHEP {\bf 0706}, 064 (2007)
  [arXiv:0705.0303 [hep-th]].

\bibitem{Alday:2007mf}
  L.~F.~Alday and J.~M.~Maldacena,
  JHEP {\bf 0711}, 019 (2007)
  [arXiv:0708.0672 [hep-th]].

\bibitem{Catani:2000ef}
  S.~Catani, S.~Dittmaier and Z.~Trocsanyi,
  Phys.\ Lett.\  B {\bf 500}, 149 (2001)
  [arXiv:hep-ph/0011222].
  
\bibitem{Penin:2005eh}
  A.~A.~Penin,
  Nucl.\ Phys.\  B {\bf 734}, 185 (2006)
  [arXiv:hep-ph/0508127].

\bibitem{Mitov:2006xs}
  A.~Mitov and S.~Moch,
  JHEP {\bf 0705}, 001 (2007)
  [arXiv:hep-ph/0612149].

\bibitem{Becher:2007cu}
  T.~Becher and K.~Melnikov,
  JHEP {\bf 0706}, 084 (2007)
  [arXiv:0704.3582 [hep-ph]].
  
\bibitem{Penin:2005kf}
  A.~A.~Penin,
  Phys.\ Rev.\ Lett.\  {\bf 95}, 010408 (2005)
  [arXiv:hep-ph/0501120].
  
\bibitem{Czakon:2007ej}
  M.~Czakon, A.~Mitov and S.~Moch,
  Phys.\ Lett.\  B {\bf 651}, 147 (2007)
  [arXiv:0705.1975 [hep-ph]].
  
\bibitem{Czakon:2007wk}
  M.~Czakon, A.~Mitov and S.~Moch,
  Nucl.\ Phys.\  B {\bf 798}, 210 (2008)
  [arXiv:0707.4139 [hep-ph]].

\bibitem{Mitov:2009sv}
  A.~Mitov, G.~Sterman and I.~Sung,
  Phys.\ Rev.\  D {\bf 79}, 094015 (2009)
  [arXiv:0903.3241 [hep-ph]].

\bibitem{Neubert:1993mb}
For a review, see:
  M.~Neubert,
  Phys.\ Rept.\  {\bf 245}, 259 (1994)
  [arXiv:hep-ph/9306320].

\bibitem{Catani:1996jh}
  S.~Catani and M.~H.~Seymour,
  Phys.\ Lett.\  B {\bf 378}, 287 (1996)
  [arXiv:hep-ph/9602277].

\bibitem{Catani:1996vz}
  S.~Catani and M.~H.~Seymour,
  Nucl.\ Phys.\  B {\bf 485}, 291 (1997)
  [Erratum-ibid.\  B {\bf 510}, 503 (1998)]
  [arXiv:hep-ph/9605323].

\bibitem{Becher:2003kh}
  T.~Becher, R.~J.~Hill, B.~O.~Lange and M.~Neubert,
  Phys.\ Rev.\  D {\bf 69}, 034013 (2004)
  [arXiv:hep-ph/0309227].

\bibitem{Bosch:2003fc}
  S.~W.~Bosch, R.~J.~Hill, B.~O.~Lange and M.~Neubert,
  Phys.\ Rev.\  D {\bf 67}, 094014 (2003)
  [arXiv:hep-ph/0301123].

\bibitem{Neubert:2004dd}
  M.~Neubert,
  Eur.\ Phys.\ J.\  C {\bf 40}, 165 (2005)
  [arXiv:hep-ph/0408179].

\bibitem{Falk:1990yz}
  A.~F.~Falk, H.~Georgi, B.~Grinstein and M.~B.~Wise,
  Nucl.\ Phys.\  B {\bf 343}, 1 (1990).

\bibitem{Korchemsky:1987wg}
  G.~P.~Korchemsky and A.~V.~Radyushkin,
  Nucl.\ Phys.\ B {\bf 283}, 342 (1987).

\bibitem{Korchemsky:1991zp}
  G.~P.~Korchemsky and A.~V.~Radyushkin,
  Phys.\ Lett.\  B {\bf 279}, 359 (1992)
  [arXiv:hep-ph/9203222].

\bibitem{Kidonakis:2009ev}
  N.~Kidonakis,
  Phys.\ Rev.\ Lett.\  {\bf 102}, 232003 (2009)
  [arXiv:0903.2561 [hep-ph]].

\bibitem{Steinhauser:2002rq}
For a review, see:
  M.~Steinhauser,
  Phys.\ Rept.\  {\bf 364}, 247 (2002)
  [arXiv:hep-ph/0201075].

\bibitem{Bernreuther:2004ih}
  W.~Bernreuther, R.~Bonciani, T.~Gehrmann, R.~Heinesch, T.~Leineweber, P.~Mastrolia and E.~Remiddi,
  Nucl.\ Phys.\  B {\bf 706}, 245 (2005)
  [arXiv:hep-ph/0406046].

\bibitem{Bernreuther:2005gq}
  W.~Bernreuther, R.~Bonciani, T.~Gehrmann, R.~Heinesch, T.~Leineweber, P.~Mastrolia and E.~Remiddi,
  Phys.\ Rev.\ Lett.\  {\bf 95}, 261802 (2005)
  [arXiv:hep-ph/0509341].
  
\bibitem{Bernreuther:2005rw}
  W.~Bernreuther, R.~Bonciani, T.~Gehrmann, R.~Heinesch, T.~Leineweber and E.~Remiddi,
  Nucl.\ Phys.\  B {\bf 723}, 91 (2005)
  [arXiv:hep-ph/0504190].

\bibitem{Georgi:1990um}
  H.~Georgi,
  Phys.\ Lett.\  B {\bf 240}, 447 (1990).
  
\bibitem{Catani:2002hc}
  S.~Catani, S.~Dittmaier, M.~H.~Seymour and Z.~Trocsanyi,
  Nucl.\ Phys.\  B {\bf 627}, 189 (2002)
  [arXiv:hep-ph/0201036].
  
\bibitem{Czakon:2008zk}
  M.~Czakon,
  Phys.\ Lett.\  B {\bf 664}, 307 (2008)
  [arXiv:0803.1400 [hep-ph]].

\bibitem{Ferroglia:2009ep}
  A.~Ferroglia, M.~Neubert, B.~D.~Pecjak and L.~L.~Yang,
  arXiv:0907.4791 [hep-ph].

\bibitem{Ferroglia:2009ii}
  A.~Ferroglia, M.~Neubert, B.~D.~Pecjak and L.~L.~Yang,
  arXiv:0908.3676 [hep-ph].

\end{thebibliography}
\end{document}